# Study of temporal and spectral characteristics of the X-ray emission from solar flares

Veena Choithani[1], Rajmal Jain[1,*], Arun K. Awasthi[2], Geetanjali Singh[1], Sneha Chaudhari[1] and Som Kumar Sharma[3]

[1] Dept. of Physics, Sh. M. M. Patel Institute of Sciences & Research, Kadi Sarva Vishwavidyalaya, Gandhinagar 382015, India; *profrajmaljain9@gmail.com

[2] CAS Key Laboratory of Geospace Environment, Department of Geophysics and Planetary Sciences, University of Science and Technology of China (USTC), Hefei, 230026, China

[3] Physical Research Laboratory, Navrangpura, Ahmedabad

**Abstract** Temporal and spectral characteristics of X-ray emission from 60 flares of intensity ≥C class observed by Solar X-ray Spectrometer (SOXS) during 2003-2011 are presented. We analyse the X-ray emission observed in four and three energy bands by the Si and CZT detectors, respectively. The number of peaks in the intensity profile of the flares varies between 1 and 3. We find moderate correlation (R≃0.2) between the rise time and the peak flux of the first peak of the flare irrespective to energy band, which is indicative of its energy-independent nature. Moreover, magnetic field complexity of the flaring region is found to be anti-correlated (R=0.61) with the rise time of the flares while positively correlated (R=0.28) with the peak flux of the flare. The time delay between the peak of the X-ray emission in a given energy band and that in the 25-30 keV decreases with increasing energy suggesting conduction cooling to be dominant in the lower energies. Analysis of 340 spectra from 14 flares reveals that the peak of Differential Emission Measure (DEM) evolution delays by 60-360 s relative to that of the temperature, and this time delay is inversely proportional to the peak flux of the flare. We conclude that temporal and intensity characteristics of flares are energy dependent as well as magnetic field configuration of the active region.

**Key words:** Sun: X-rays, gamma rays — Sun: magnetic fields — Sun: flares — Sun: Corona

## 1 INTRODUCTION

Solar flares are understood to be a consequence of the magnetic reconnection process occurring in the solar corona followed by the release of magnetic energy which is stored a-priori (Jain 1983; Mirzoian 1984; Haisch et al. 1991; Jain 1993; Garcia Alvarez 2000; Jain et al. 2000; Uddin et al. 2004; Chandra et al. 2006; Jain et al. 2008, 2011a; Choudhary et al. 2013; Awasthi et al. 2014, 2018a). Earlier, Dennis (1985) carried out detailed study of over 7000 solar flares observed by the SMM to understand the energy release and particle transport processes that lead to the high-energy X-ray aspects of solar flares. He discovered 152-158 day periodicity in various aspects of solar activity including the rate of occurrence of hard X-ray and gamma-ray flares. A typical high-intensity class flare may register enhancement covering the whole electromagnetic spectrum. However, X-ray spectra from a typical large solar flare is dominated by soft X-ray line and thermal (free-free) bremsstrahlung emission at $\epsilon$=1-20 keV, and collisional bremsstrahlung of non-thermal electrons at $\epsilon$=20-1000 keV (Jain et al. 2000, 2005, 2008, 2011b; Bhatt et al. 2013). In-depth investigation of the soft and hard X-ray flux prior to and during a flare provides a diagnostics of the temporal and spectral X-ray characteristics of high-temperature flare plasma (Jain 1993; Jain et al. 2000; Aggarwal et al. 2008; Jain et al. 2011b; Chowdhury et al. 2013; Awasthi et al. 2016, 2018b). Jain et al. (2008, 2011b) have shown that the X-ray bursts associated with solar flares above 25 keV are mostly associated with the non-thermal bremsstrahlung while the X-ray emission between 10 and 25 keV in principle includes both thermal and non-thermal contributions. Both the aforementioned types of X-ray emissions are present during solar flares and a clear demarcation between the two processes provides quantitative estimation of detailed and complete energetic of solar flares. In this regard, temporal and spectral mode investigation of X-ray emission during solar flares is of immense value.

In particular, statistical as well as case studies by investigating radiated intensity in X-ray waveband during solar flares has been made by several researchers in the past to determine temporal characteristics of the flare including rise time, energy-dependent delay in the peak emission etc. With the use of automated algorithm for the detection of flare onset and peak times, a statistical study made by Zhang & Liu (2015) for the GOES flares observed in the years 1980-2013 resulted in the fact that the rise time of flares is about half of the decay time. Moreover, they found that although it is shorter in high-energy channels than in the low-energy band, it increases with the flux increased from the start time to the peak time of a flare. Further, Su et al. (2006); Yang et al. (2009) in their study concluded a weak correlation ( 0.4) between the peak count rates and the characteristic times like rise times, decay times, or flare durations. On the contrary, the correlation between the rise times and decay times is derived to be strong (0.76) in their study. Kubo & Akioka (2004) suggested that the SXR fluence of a flare is a better indicator than peak intensity of the SXR because of the time integral of X-ray flux, which is related to the total energy released by the flare. Therefore, although the aforementioned studies presented exhaustive analysis of temporal characteristics of the flare, a detailed study of the dependence of temporal characteristics on energy channel is poor.

The energetic properties of the flares are revealed by the detailed analysis of the spectra. It is well established that the solar corona emits strongly in the EUV and X-ray part of the spectrum through both emission lines and continuum in general and in particular during flares. Particularly in the regime of X-ray waveband, the spectrum contains a multitude of emission from lines, two-photon, free-free and free-bound continuum. To fully understand the origin and the conditions of the plasma emitting such radiation, it is necessary to model the spectrum in this regime. In order to do so, parameters such as the energy levels, transitions, radiative transfer probabilities and excitation rates must be well understood for each individual line as well as continuum process (Phillips 2004; Jain et al. 2005, 2006, 2008, 2010). Currently for line emission analysis we make use of the CHIANTI atomic physics package (Dere et al. 1997; Landi et al.

2013). However, in this paper we do not focus on the analysis of line emission rather we concentrate on the continuum emission by thermal (soft X-ray) and non-thermal X-ray (hard X-ray) bremsstrahlung process (Jain et al. 2008, 2011b). In hot coronal plasma, free electrons and ions can suffer multiple interactions. The most frequent of these is when a free electron is scattered in the Coulomb field of an ion (Z). The scattering is such that the electron remains free after the interaction. This is what gives rise to continuum emission also known as bremsstrahlung radiation. Thick target bremsstrahlung occurs when electrons are accelerated to high energies in collisionless plasma and get stopped when they interact with thermal plasma such as the chromospheres (Jain et al. 2005, 2008). Hence in order to understand the energetic characteristics of the flare plasma we must exploit the flare spectra in greater detail. Thus the aim of this paper is to understand the temporal and spectral characteristics of solar flares which in turn may improve our current knowledge on the physical processes that occur during conductive and radiative cooling phases of the flare.

We consider 60 flares observed by the Solar X-ray spectrometer (SOXS) mission (Jain et al. 2005) to study the temporal characteristics. Out of which we consider 14 flares for detailed spectra analysis, which enables us to measure the plasma parameters viz. temperature, DEM, spectral nature (thermal versus non-thermal) etc. Study of temporal variation of these spectral characteristics is utmost essential to probe how they vary over time and with magnitude of the flare. Therefore in this paper, we investigate the temporal and spectral behaviour of flare plasma. In section 2, we present brief description of the instrument and observations. Section 3 deals with the temporal characteristics of the flare plasma, while section 4 consists of spectral evolution of flare plasma parameters. In section 5, we discuss and conclude the paper.

## 2 OBSERVATIONS

We investigate the dependence of temporal characteristics of the flare plasma viz. rise time and peak flux as a function of the emitted energy. In this regard, we analyse full disk integrated X-ray emission from 30 M-class and 30 C-class flares recorded by the Solar X-ray spectrometer (SOXS) instrument during year 2003-2011.

SOXS was flown onboard GSAT-2 satellite on 08 May 2003. SOXS was a high spectral and temporal resolution X-ray spectrometer to study the disk-integrated flux from the Sun in 4 keV-10 MeV energy range with a time resolution of 10-100 ms using semiconductor devices and Phoswitch Scintillation Detector (Jain et al. 2000). SOXS had two types of detector modules viz. SOXS Low Energy Detector (SLD) Module and SOXS High Energy Detector (SHD) Module. The SLD module consisted of state-of-the-art semiconductor devices viz. Silicon-P-intrinsic-N (PIN) detector for the low energy X-rays (4 to 25 keV) and Cadmium-Zinc-Telluride (CZT) detector for soft to medium X-rays in the energy range 4 to 56 keV. On the other hand, SHD module consisted of large Phoswitch Scintillation detector for high energy X-rays and Gamma rays (15 keV to 10 MeV). The SLD and SHD both modules were designed to work independently so as to observe the disk integrated radiative flux from the sun to enable us to study the mechanisms of energy release by solar flares (Jain et al. 2005).The details of instrumentation, calibration, observations, and temporal and energy resolution of the Si and CZT detectors have been presented earlier by (Jain et al. 2005, 2006, 2008, 2011b).

## 2.1 Data Set

Physical properties of the selected 60 flares are presented in the Table I. The start time, peak time, and peak flux for each flare are measured in the energy bands: 5-7 keV, 7-9 keV, 9-11 keV and 11-15 keV of Si detector, and 15-20 keV, 20-25 keV and 25-30 keV of CZT detector.

Table 1: Physical Properties of solar flares under current investigation

| Sr. No. | Date | 5-7 keV | | | GOES Class | Location | NOAA Number |
| --- | --- | --- | --- | --- | --- | --- | --- |
| | | Start Time (U.T.) | Peak Time (U.T.) | End Time (U.T.) | | | |
| 1 | 2003 Jul. 30 | 4:07:12 | 4:09:39 | 4:31:24 | M 2.5 | N14W55 | 10422 |
| 2 | 2003 Nov. 13 | 4:58:47 | 5:02:11 | 5:14:59 | M 1.6 | N01E90 | 10501 |
| 3 | 2003 Nov. 19 | 3:57:02 | 4:00:32 | 4:23:47 | M 1.7 | N01E06 | 10501 |
| 4 | 2004 Jan. 06 | 6:16:32 | 6:22:20 | 7:05:50 | M 5.8 | N05E90 | 10537 |
| 5 | 2004 Jan. 07 | 3:56:00 | 4:01:12 | 4:41:57 | M 4.5 | N02E82 | 10537 |
| 6 | 2004 Jan. 08 | 4:56:00 | 5:04:57 | 5:35:21 | M 1.3 | N01E64 | 10537 |
| 7 | 2004 Jan. 10 | 4:18:06 | 4:20:12 | 4:45:12 | C 7.3 | S13W32 | 10536 |
| 8 | 2004 Jan. 10 | 5:10:03 | 5:13:00 | 5:29:39 | C 7.7 | S13W32 | 10536 |
| 9 | 2004 Jan. 19 | 4:51:29 | 4:55:32 | 5:08:29 | M 1.0 | S07W75 | 10500 |
| 10 | 2004 Feb. 09 | 5:51:26 | 5:54:41 | 6:30:47 | C 5.5 | S08E48 | 10554 |
| 11 | 2004 Mar. 22 | 6:09:16 | 6:16:04 | 6:48:04 | C 8.6 | S04W03 | 10574 |
| 12 | 2004 Mar. 25 | 4:29:12 | 4:34:00 | 5:13:30 | M 2.3 | N12E82 | 10582 |
| 13 | 2004 Apr. 05 | 5:35:50 | 5:43:26 | 6:30:05 | M 1.7 | S18E35 | 10588 |
| 14 | 2004 Apr. 11 | 3:58:14 | 4:10:41 | 5:06:32 | C 9.6 | S14W47 | 10588 |
| 15 | 2004 Apr. 25 | 5:28:36 | 5:33:27 | 6:11:15 | M 2.2 | N13E38 | 10599 |
| 16 | 2004 Jul. 12 | 4:28:24 | 4:32:12 | 5:02:45 | C 5.5 | S11E84 | 10649 |
| 17 | 2004 Jul. 13 | 5:26:31 | 5:29:37 | 5:34:25 | C 6.7 | N15W49 | 10646 |
| 18 | 2004 Jul.14 | 5:17:13 | 5:18:55 | 5:33:22 | M 6.2 | N12W62 | 10646 |
| 19 | 2004 Jul. 21 | 5:11:59 | 5:17:53 | 5:32:47 | C 8.9 | N05E24 | 10652 |
| 20 | 2004 Aug. 14 | 4:11:51 | 4:14:30 | 4:54:45 | M 7.4 | S11W28 | 10656 |
| 21 | 2004 Aug. 14 | 5:36:18 | 5:40:42 | 6:03:57 | M 2.4 | S11W28 | 10656 |
| 22 | 2004 Aug. 17 | 4:59:37 | 5:02:46 | 5:43:22 | M 1.1 | S12W66 | 10656 |
| 23 | 2004 Aug. 31 | 5:27:50 | 5:35:35 | 6:01:53 | M 1.4 | N06W84 | 10663 |
| 24 | 2004 Oct. 28 | 6:03:27 | 6:04:54 | 6:21:27 | C 3.8 | S19E72 | 10693 |
| 25 | 2004 Oct. 30 | 6:10:09 | 6:13:54 | 6:24:21 | M 4.2 | N14W21 | 10691 |
| 26 | 2004 Oct. 31 | 5:25:00 | 5:30:00 | 5:53:48 | M 2.3 | N13W24 | 10691 |
| 27 | 2004 Nov. 07 | 4:12:36 | 4:15:03 | 4:35:24 | C 5.1 | N10W08 | 10696 |

Table 1: Table-1 Continued

| Sr. No. | Date | 5-7 keV | | | GOES Class | Location | NOAA Number |
|---|---|---|---|---|---|---|---|
| | | Start Time (U.T.) | Peak Time (U.T.) | End Time (U.T.) | | | |
| 28 | 2004 Nov. 07 | 4:45:39 | 4:47:39 | 5:05:36 | C 5.5 | N10W08 | 10696 |
| 29 | 2004 Nov. 19 | 5:05:44 | 5:09:59 | 5:36:41 | C 4.9 | N05W90 | 10700 |
| 30 | 2005 Jan. 15 | 4:10:10 | 4:14:04 | 4:24:25 | M 8.4 | N14E06 | 10720 |
| 31 | 2005 Jan. 15 | 4:28:43 | 4:30:43 | 4:44:58 | M 1.3 | N14E06 | 10720 |
| 32 | 2005 Jan. 21 | 4:19:09 | 4:25:51 | 4:55:06 | C 6.3 | N19W69 | 10720 |
| 33 | 2005 Jul. 01 | 4:57:55 | 5:00:52 | 5:15:10 | C 5.3 | N14E83 | 10786 |
| 34 | 2005 Jul. 03 | 4:48:36 | 4:51:27 | 5:13:27 | C 4.7 | S09W39 | 10787 |
| 35 | 2005 Jul. 27 | 4:35:35 | 4:47:23 | 5:28:47 | M 3.7 | N13W05 | 10792 |
| 36 | 2005 Jul. 30 | 5:06:32 | 5:13:02 | 5:43:41 | C 9.4 | N12E54 | 10792 |
| 37 | 2005 Aug. 03 | 4:55:30 | 5:02:21 | 5:28:30 | M 3.5 | S11E36 | 10794 |
| 38 | 2005 Aug. 25 | 4:34:03 | 4:39:24 | 5:07:39 | M 6.4 | N07E78 | 10803 |
| 39 | 2005 Sep. 09 | 5:31:30 | 5:39:15 | 6:09:33 | M 6.2 | S10E66 | 10808 |
| 40 | 2005 Sep. 12 | 4:46:54 | 5:00:57 | 5:33:42 | M 1.5 | N10E00 | 10809 |
| 41 | 2005 Sep.17 | 5:55:30 | 6:01:21 | 6:28:30 | M 9.8 | S10W39 | 10808 |
| 42 | 2005 Nov. 14 | 3:57:03 | 3:58:51 | 4:11:51 | M 2.6 | S07E68 | 10822 |
| 43 | 2005 Nov. 14 | 4:17:03 | 4:21:48 | 4:57:00 | C 7.3 | S07E68 | 10822 |
| 44 | 2005 Dec. 01 | 4:51:50 | 4:54:05 | 5:18:56 | C 5.7 | S04E29 | 10826 |
| 45 | 2006 Apr. 06 | 5:24:45 | 5:32:48 | 6:02:09 | M 1.4 | S07W54 | 10865 |
| 46 | 2006 Dec. 05 | 5:03:15 | 5:07:30 | 5:25:36 | C 4.2 | S07E90 | 10930 |
| 47 | 2006 Dec 07 | 4:29:47 | 4:38:23 | 5:05:08 | C 6.1 | S08E55 | 10930 |
| 48 | 2010 Feb. 08 | 5:14:15 | 5:21:57 | 5:33:42 | C 8.6 | N21W01 | 11045 |
| 49 | 2010 Oct. 31 | 4:27:31 | 4:28:46 | 4:47:19 | C 5.7 | N20W86 | 11117 |
| 50 | 2010 Nov. 06 | 4:39:39 | 4:46:57 | 5:15:36 | C 4.5 | S18E44 | 11121 |
| 51 | 2011 Feb. 14 | 4:27:46 | 4:37:46 | 5:49:25 | C 8.3 | S20W17 | 11158 |
| 52 | 2011 Feb. 15 | 4:28:37 | 4:30:13 | 4:37:46 | C 4.8 | S20W15 | 11158 |
| 53 | 2011 Feb. 16 | 5:41:48 | 5:45:03 | 5:59:27 | C 5.9 | S21W41 | 11158 |
| 54 | 2011 Feb. 18 | 4:45:22 | 4:49:55 | 5:19:49 | C 4.0 | N11E38 | 11161 |
| 55 | 2011 Mar. 07 | 5:01:42 | 5:08:12 | 5:26:27 | M 1.2 | N23W47 | 11164 |
| 56 | 2011 Mar. 09 | 4:58:34 | 5:02:22 | 5:07:28 | C 8.0 | N11W15 | 11166 |
| 57 | 2011 Mar. 11 | 4:28:59 | 4:31:59 | 4:37:38 | C 5.5 | N08W32 | 11166 |
| 58 | 2011 Mar. 12 | 4:35:03 | 4:39:51 | 4:51:42 | M 1.2 | N05W36 | 11166 |
| 59 | 2011 Apr. 14 | 5:23:08 | 5:24:59 | 5:34:50 | C 4.9 | N15E69 | 11193 |
| 60 | 2011 Apr. 22 | 4:36:55 | 4:41:16 | 5:17:49 | M 1.8 | S17E42 | 11195 |

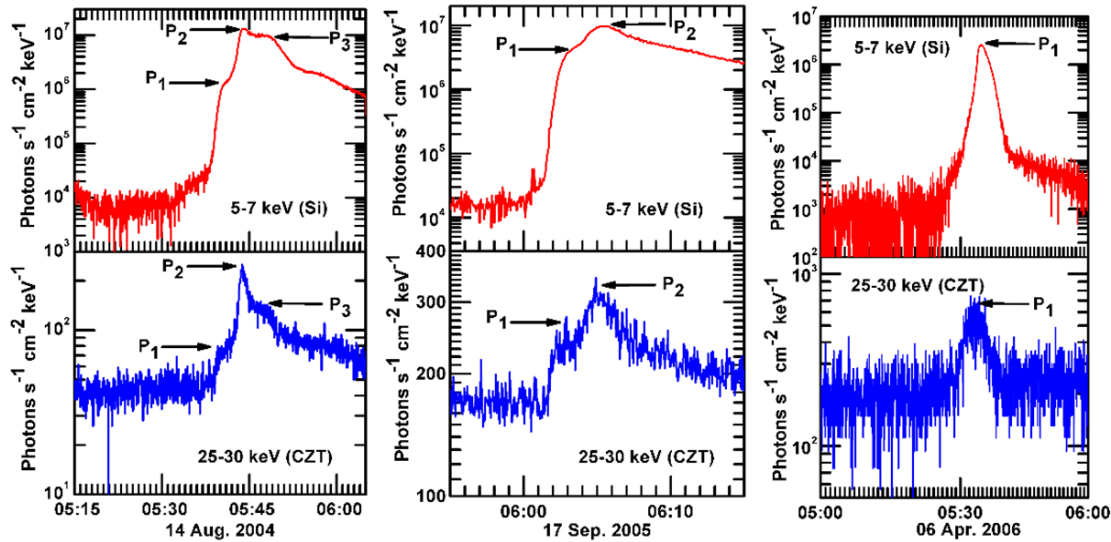

Fig. 1: Temporal evolution of the X-ray photon flux in 5-7 keV (red) and 25-30 keV (blue) for the solar flares observed on 14 Aug2004(left panel), 17 September 2005 (middle panel) and 06 April 2006 (right panel).

Although the temporal characteristics of the flares in only 5-7 keV energy band are listed in Table 1, for completeness we provide the characteristics deduced in rest of the aforementioned energy bands in the Appendix Table 1.

## 3 TEMPORAL CHARACTERISTICS OF SOLAR FLARES

In order to probe the temporal characteristics of the emission during flare, X-ray intensity recorded in seven energy channels, four from Si detector viz. 5-7 keV, 7-9 keV, 9-11 keV and 11-15 keV and three from CZT detector viz.15-20 keV, 20-25 keV and 25-30 keV (hard X-ray) has been analyzed for 60 flares. Temporal cadence of the observations used for current time characteristics is 3s (Jain et al. 2005). Temporal and spectral mode data in count flux is converted into detector independent quantity namely, the photon flux, for the current analysis employing response matrix for Si and CZT detectors. The temporal evolution of photon flux in all seven energy channels for all 60 flares has been studied in greater detail. However, in order to demonstrate, we show in Figure 1 the temporal evolution in the 5-7 (soft) and 25-30 keV (hard) energy bands for the flares observed on 14 August 2004 (left panel), 17 September 2005 (middle panel) and 06 April 2006 (right panel).

In order to determine the temporal characteristics of the flare such as start time, peak time, peak flux, and end time, we prepared a semi-automatic algorithm in Interactive Data Language (IDL). After plotting the intensity profile of a flare in various energy bands, we manually selected the time duration corresponding to the pre-flare background which is then provided as an input to derive the averaged background value as well as its standard deviation ($\sigma$). In a given energy band, the start time of the flare is considered to be the time when the flux reaches $\geq 3\sigma$. Next, in order to determine the peak flux and peak time, we select two points on the intensity profile in a way so that they cover the time instance at which the flux attained its maximum. The algorithm makes use of the time duration selected above to deduce the peak count. This peak intensity, after subtracting the background, is considered as the peak flux while the time at which the intensity attained its maximum is termed as the peak time. The aforementioned procedure is repeated for each individual peak for the flare cases where the peaks are more than one. However, the start time is considered to be common for all the peaks. Rise time (t1) is defined as start time subtracted from peak one. Similarly, rise time for peak 2 and peak 3 viz. t2 and t3 are obtained by subtracting start time from their respective peak time. A schematic diagram representing the definition of rise times considered in this investigation is shown in Figure 2.

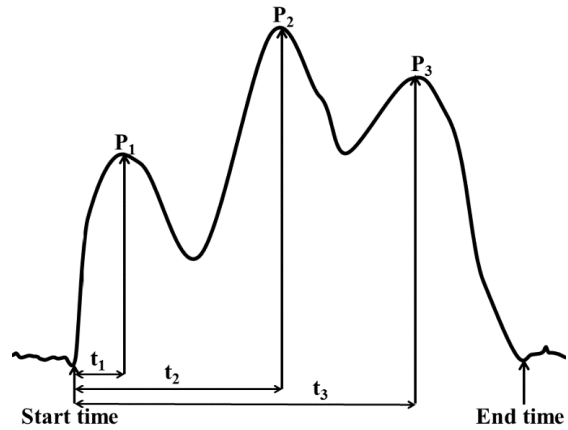

Fig. 2: Schematics of flare evolution and rise time.

Investigation of temporal evolution of the emission recorded for 60 flares comprised of 29 flares showing (13 M Class, 16 C Class) single peak, 23 flares showing double peak (14 M Class and 9 C Class), while rest of the 8 flares show the evolution consisting of 3 steps (3 M Class and 5 C Class). In Figure 1 we show the representative cases depicting the evolution of aforementioned different types. As following, we present the comprehensive statistics of start time, peak time, rise time of the flare and their relationship with peak flux in soft and hard X-rays energy bands. A histogram of the time delay between start time and peak times of SXR (5-7 keV) and HXR (25-30 keV) emission is prepared as shown in Figure 3. It may be noted that the SXR enhancement tends to be detected earlier than the HXR emission. On the other hand, the maximum phase in the SXR emission is delayed from the HXR peak with a mean delay for peak 1 is estimated to be 48 s.

Next, we estimate the rise time for all the peaks i.e. t1, t2 and t3 corresponding to various energy bands within 5-30 keV. The distribution of rise time corresponding to peak 1 is shown in the form of histogram for all the energies in Figure 4. From this distribution, we note that for most of the flares, rise time of the higher energy band is lower than that of lower energy bands.

We further investigate the correlation of rise time with energy as plotted in Figure 5. We deduced the mean rise time of the flare for first peak to be 433s and 132 s in 5-7 keV and 25-30 keV, respectively. Further, from our temporal analysis, we find that mean rise time in soft X-ray regime (5-7 keV) remains >200 s, while in hard X-ray regime (25-30 keV) the same attains the values <200 s. Interestingly, this investigation revealed that rise time decreases as a function of energy. Jain et al. (2011b) found the X-ray light curves in different energy bands to peak at different time instances, with the intensity profile corresponding to the highest energy to peak the earliest when compared with that in lower energies. Moreover, Aschwanden (2007), modelled energy-dependent delay in the X-ray intensity profiles, observed from *RHESSI*, in terms of exponentially decaying cooling. This implies that longer cooling time scale will cause more delay in attaining peak in case of low energies which explains the relationship of rise time with the energy that revealed in our investigation.

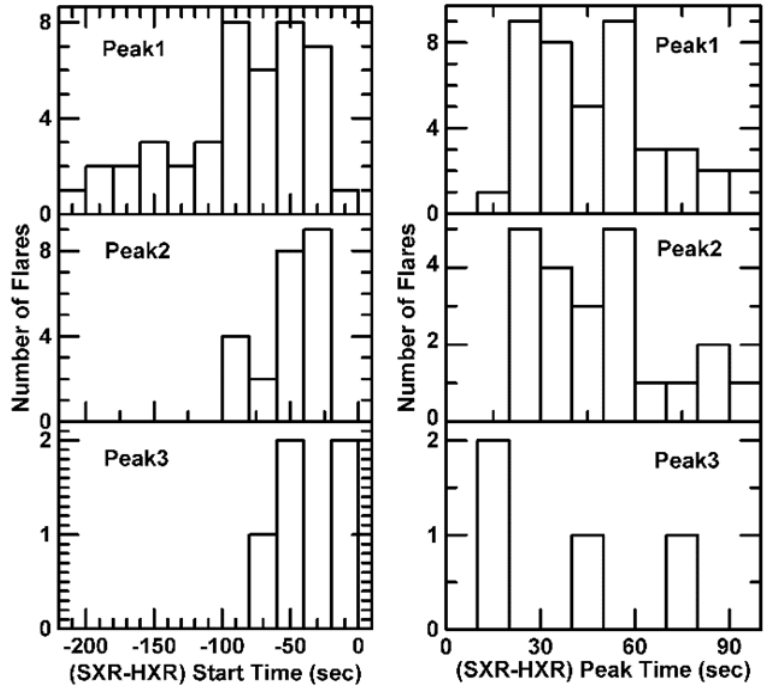

Fig. 3: Histogram of the delay between the onset (left) and peak times (right) of the SXR **(5-7 keV)** and HXR **(25-30 keV)** emission for all the flare with one peak (top panel), two peaks (middle panel) and three peaks (bottom panel).

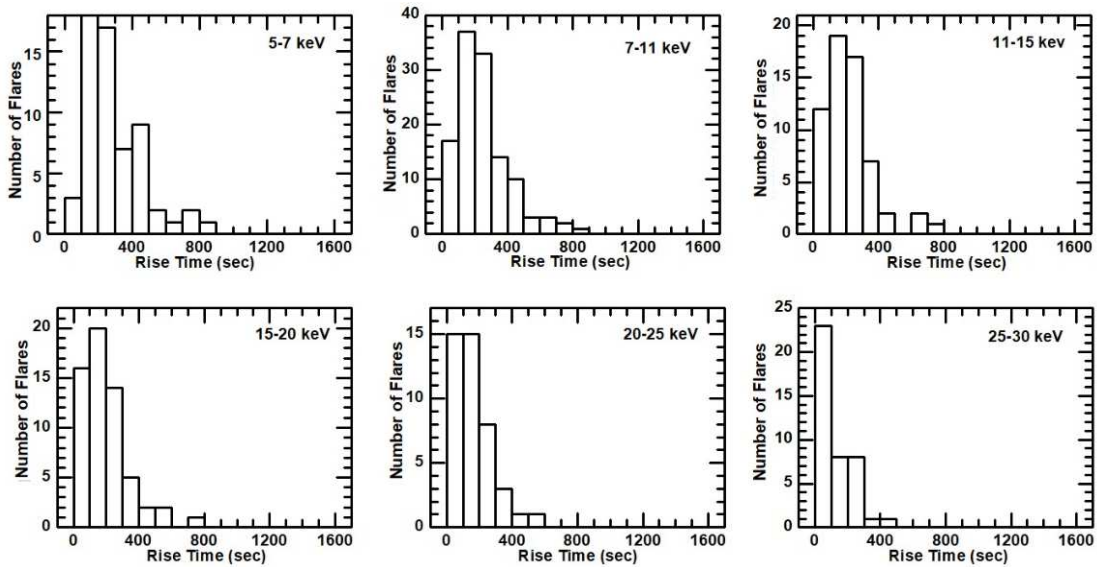

Fig. 4: Histograms of rise time for first peak of the flares calculated in energy bands ranging from 5-30 keV.

Next, we derive the energy dependence of the relationship of the rise time with the flare intensity. As shown in Figure 6, we plot the rise time as a function of the peak flux in seven energy bands covering 5-30 keV (viz. 5-7, 7-9, 9-11, 11-15, 15-20, 20-25 and 25-30 keV).

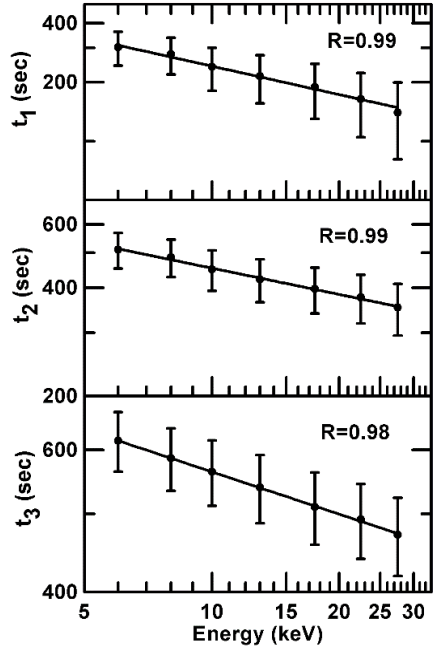

Fig. 5: Variation of the flare mean rise time as a function of energy bands covering 5-30 keV for flares with different number of peaks where top panel demonstrates the relationship corresponding to the first peak while the middle and bottom panels show that for peak 2 and 3, respectively.

It may be noted from the Figure 6(a), (b), (c) that the rise time is positively, yet weakly correlated with peak flux in the energy range 5-30 keV. Further, while the aforementioned correlation, derived for first peak (t1) is unambiguous, the same corresponding to t2 and t3 is inevitably less trustworthy owing to the effect of the higher background due to pre-heated plasma. Furthermore, our investigation revealed aforesaid correlation to be independent of energy-band choice. The correlation coefficient (R) in this study remained less than 0.4 irrespective to the different peaks of the flares. Although our finding is based on a very limited data set, the results are in good agreement with that obtained in Veronig et al. (2002), who found correlation coefficient R=0.25. It may be noted that Veronig et al. (2002) employed observations in 1.55-12.4 keV (1-8 Å) of GOES, which, belong to the soft X-ray energy bands compared to the extensively large energy band coverage in our investigation. On the other hand, if we compare our mean correlation coefficient in soft X-ray bands (5-15 keV), it results to R=0.28, which is in close agreement with that obtained in Veronig et al. (2002). For peak 1 and 3, the correlation coefficient is found to be higher in the HXR energy band (25-30 keV) compared to that derived for all the other energy bands, which, however, is not the case for peak 2.

Next, we also investigate the relation of flare evolution in terms of rise time and peak flux with the magnetic complexity, $M_j$, defined earlier by Jain (1983). According to Jain (1983) complexity of the magnetic field ($M_j$) of a sunspot group may be determined from its class, polarity and field strength. Using all these information, magnetic complexity index ($M_j$) of sunspot group is determined for each day until it is visible on the disk as follows.

The magnetic complexity number Mj is defined by the following empirical relation.

$$M_j = W_j + B_j \tag{1}$$

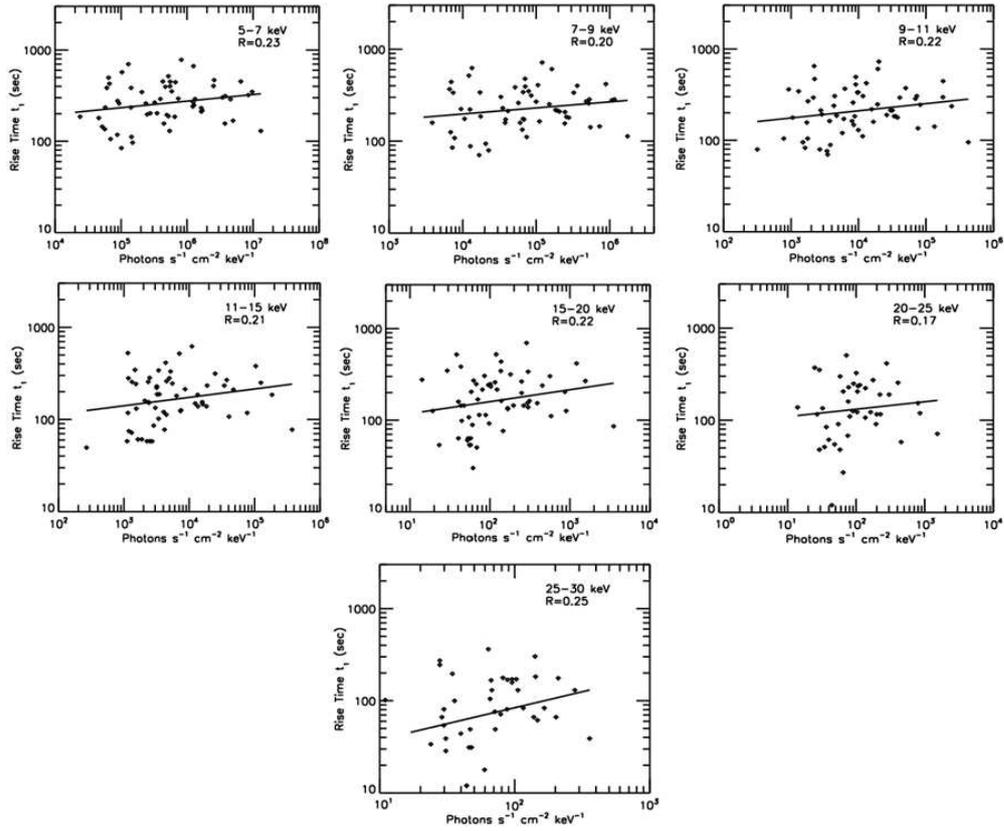

(a) Plot of rise time (t1) as a function of peak flux corresponding to energy bands covering 5-30 keV for evolution of flares.

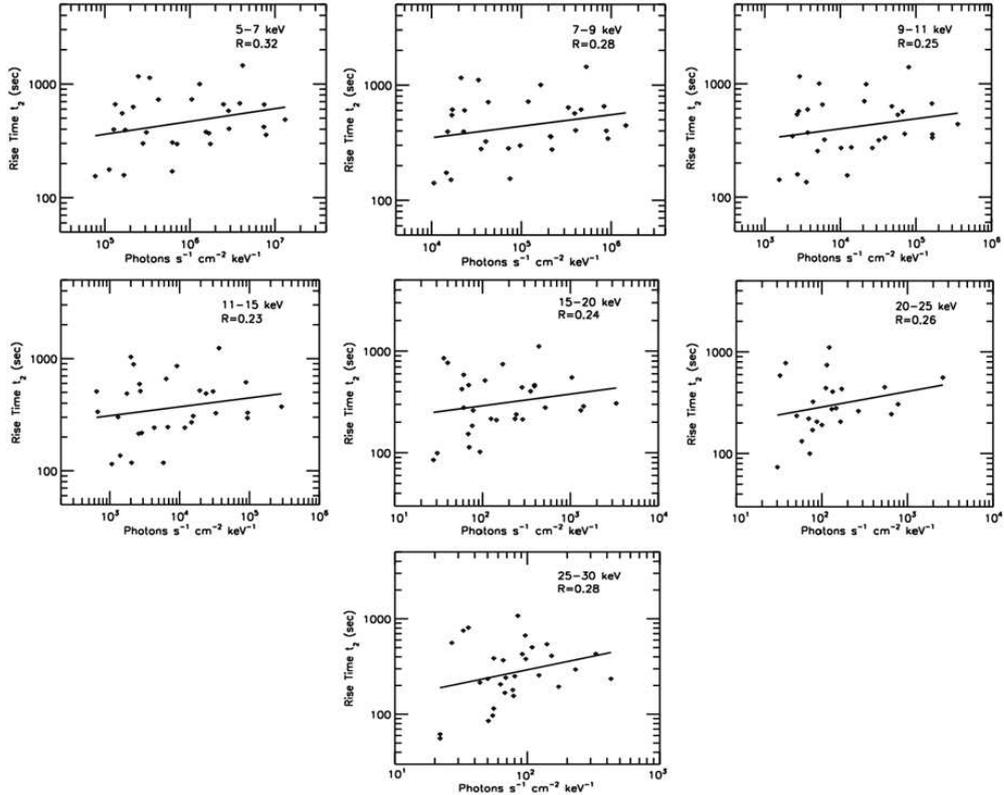

(b) Plot of rise time (t2) as a function of peak flux corresponding to energy bands covering 5-30 keV for evolution of flares.

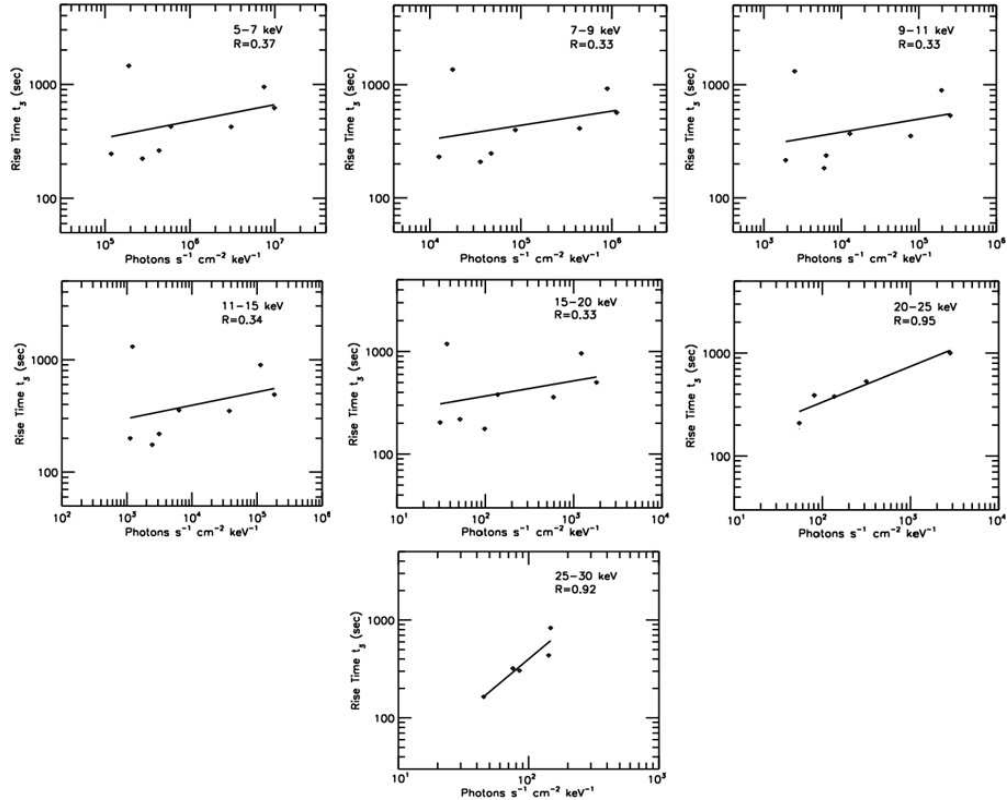

(c) Plot of rise time (t3) as a function of peak flux corresponding to energy bands covering 5-30 keV for evolution of flares.

Fig. 6: Correlation of the rise time (t1, t2 and t3 in panel (a), (b) and (c), respectively) with the peak flux.

Where, $W_j$ is code number given to Mt. Wilson class of sunspot group and $B_j$ is code number given to Boulder sunspot class. Zurich classification provides information about the sunspot class. The modified Zurich sunspot group classification is comprised of three letters generalized as 'Zpc'. In this, Z stands for the Zurich Class (denoted as A,H,B,C,D,E,F), p represents the Penumbra type of largest spot in group (represented by x, r, s, a, h, k) and c is Relative sunspot distribution or compactness of the group and represented with the letters x, o, i, and c. Each sub classification of letter is assigned with numeric number. Further, Mount Wilson magnetic classification gives information about the magnetic class (A, B, G, BG, Delta, BD, BGD). This class is also assigned with numbers (Bray & Loughhead 1964). We used these codes to find out magnetic complexity number Mj.

For example, the magnetic complexity number $M_j$, for Mc Math Region 10656 on 14th August 2004 is computed from the table "Regions of Solar Activity" being published in SGD reports, as follows:

$$W_j = BG = 3 \qquad (2)$$

$$B_j = FKC = 7 + 6 + 4 = 17 \qquad (3)$$

Substituting equation (2) and (3) in equation (1), $M_j$= 3+17 = 20

Similarly, for Mc Math Region 10808 on 17th September 2005,

$$W_j = BD = 7 \qquad (4)$$

$$B_j = EKC = 6 + 6 + 4 = 16 \qquad (5)$$

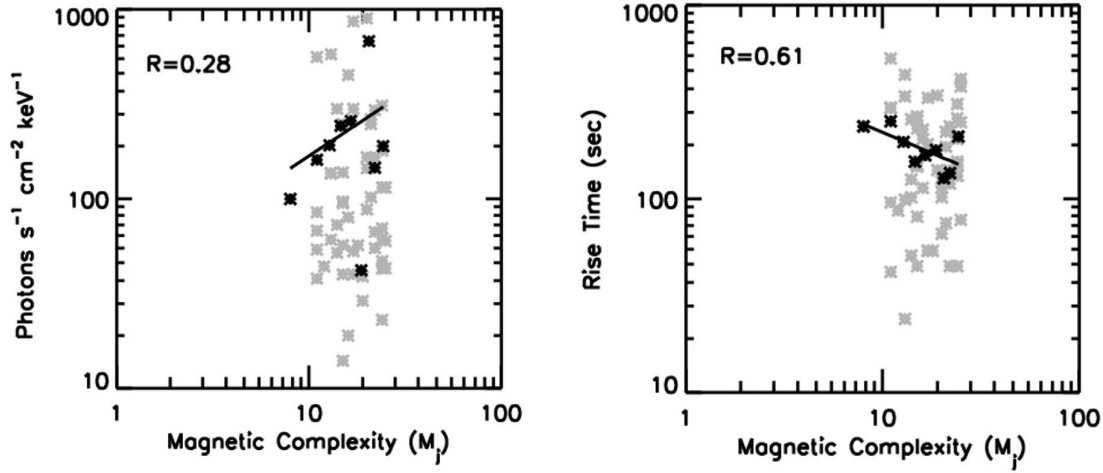

Fig. 7: Variation of peak intensity (left panel) and rise time (right panel) as a function of magnetic complexity ($M_j$).Correlation coefficient is determined by fitting the binned data points (black symbol) while the original data points are also plotted in grey for the reference.

Substituting equation (4) and (5) in equation (1), $M_j$= 7+16 = 23

Similarly, for Mc Math Region 10865 on 06th April 2006,

$$W_j = BG = 3 \qquad (6)$$

$$B_j = EKO = 6 + 6 + 2 = 14 \qquad (7)$$

Substituting equation (6) and (7) in equation (1), $M_j$= 3+14 = 17

The detailed codes for each letter (upper and lower case) have been described by Jain (1983) and presented in Table 2.

Employing above methodology, we derived $M_j$ for all 60 flares and to avoid statistical noise we carried out binning of $M_j$ at 2 point level. We find that the flare evolution in terms of rise time corresponding to the peak flux (maximum in case of multiple peak flares) to be highly correlated (R=0.61) with the magnetic complexity. On the other hand, the peak magnitude of the flux is found to be moderately correlated (R=0.28) with the magnetic complexity index, $M_j$ as shown in Figure 7. We found almost similar correlation coefficient values from the investigation of the dependence of rise time with the peak flux for the peak 1 (Figure 6a). This figure reveals unambiguously key role of magnetic field complexity in governing the magnitude and rise time of the flare.

Strong correlation (R=0.83) is found between the peak fluxes derived in the energy bands 5-7 keV (soft X-ray), and 25-30 keV (hard X-ray) as shown in Figure 8. For the same, we considered all the 60 flare events investigated in this work, and in particular, for the flares with multiple peaks, we considered the maximum of all the respective peaks within a flare for correlation study. This suggests that the energetic electrons are responsible for both the hard X-ray emission via thick-target collisional bremsstrahlung as well as the main source of heating and mass supply via chromospheric evaporation that produce soft X-ray emission (Veronig et al. 2005; Jain et al. 2008, 2011b). Thus time derivative of soft X-emission must reveal temporal evolution of hard X-ray emission known as Neupert effect. Thus our finding shown in Figure 8 supports in principle the Neupert effect, which may be further evidenced from Figure 9.

Table 2: Specific codes for parameters of Magnetic Complexity Index ($M_j$)

| Wj | | Bj | | | | | |
|---|---|---|---|---|---|---|---|
| Mt.Wilson class | Code | Zurich class (1st letter) | Code | McIntosh Classification | | | |
| | | | | 2nd letter | Code | 3rd letter | Code |
| AP | 1 | A | 1 | x | 1 | x | 1 |
| AF | 1 | H | 2 | r | 2 | o | 2 |
| BP | 2 | B | 3 | s | 3 | i | 3 |
| BF | 2 | C | 4 | a | 4 | c | 4 |
| B | 2 | D | 5 | h | 5 | | |
| BY | 3 | E | 6 | k | 6 | | |
| Y | 4 | F | 7 | | | | |
| D | 5 | | | | | | |

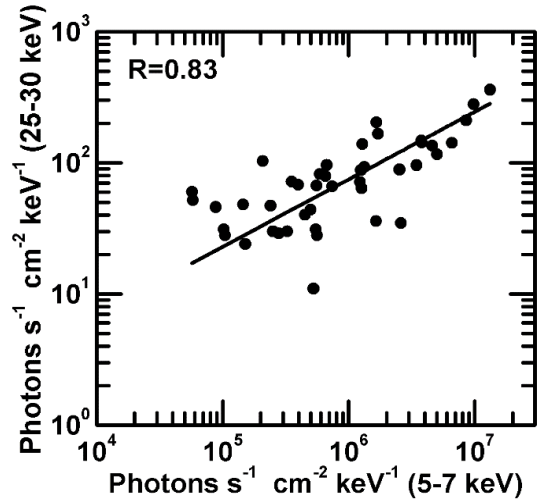

Fig. 8: Plot of hard x-ray peak flux (25-30 keV) as a function of soft x-ray peak flux (5-7 keV).

Further, we measured peak time of a given flare in the 25-30 keV energy band, reference energy band, and compared the same with the peak time in the lower energy bands, and refer the same as 'delay time' ($\Delta t$). Figure 10 shows the distribution of delay time for peak 1.

Measurement of delay time ($\Delta t$) between the peak time of a given energy band relative to the reference energy band allowed us to estimate that the delay time to be highest (54 s) for the 5-7 keV, and reducing to 10 s towards higher energy bands (20-25 keV) as shown in Figure 11.

## 4 SPECTRAL CHARACTERISTICS OF SOLAR FLARES

We use the OSPEX (Object Spectral Executive) software package inside Solar-Soft package to analyze the data. The OSPEX is an object-oriented interface for X-ray spectral analysis of solar flare data. It is a new ver-sion of the original SPEX (Spectral Executive) command line tool written by R. Schwartz in 1995. Through OSPEX, it has been made possible to load the SOXS data, available in the public domain, downloadable in the '.les (low energy spectra)' format, display the temporal evolution in the energy bands of interest, selecting and subtracting the background, and subsequently employing a combination of photon flux model on the data recorded in various time intervals of interest to describe the data by iteratively forward fitting. With this tool, we estimated the temporal evolution of the thermal and non-thermal plasma parameters by the use of the CHIANTI code (Dere et al. 1997; Landi et al. 2013) for flare plasma diagnostics with the application of various thermal, line emission, multi-thermal, and non-thermal functions (Jain et al. 2008, 2011b; Awasthi et al. 2014).

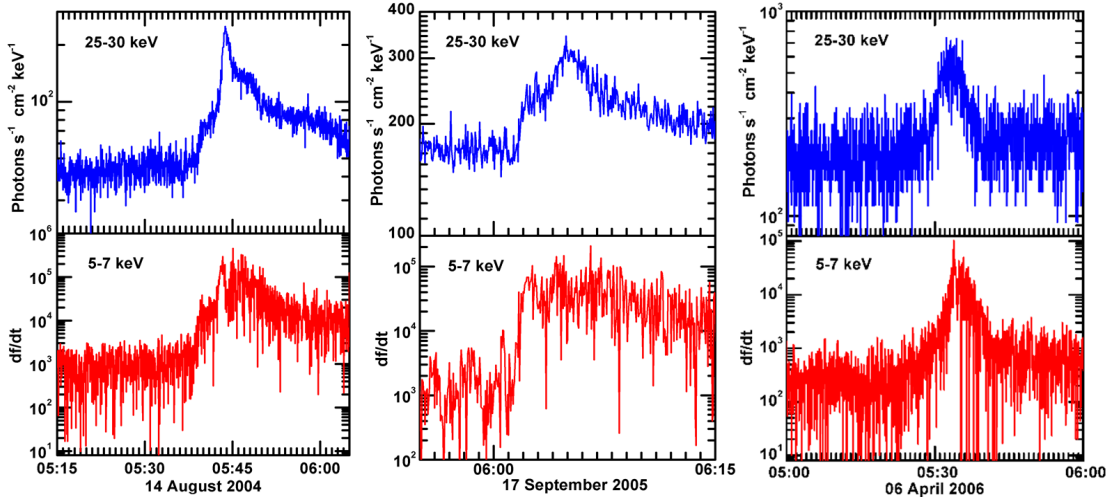

Fig. 9: Time derivative of soft X-ray emission in 5-7 keV depicts temporal evolution of the HXR emission in accordance to Neupert effect. The flares demonstrated are: 14 August 2004 (left panel), 17 September 2005 (middle panel), and 06 April 2006 (right panel).

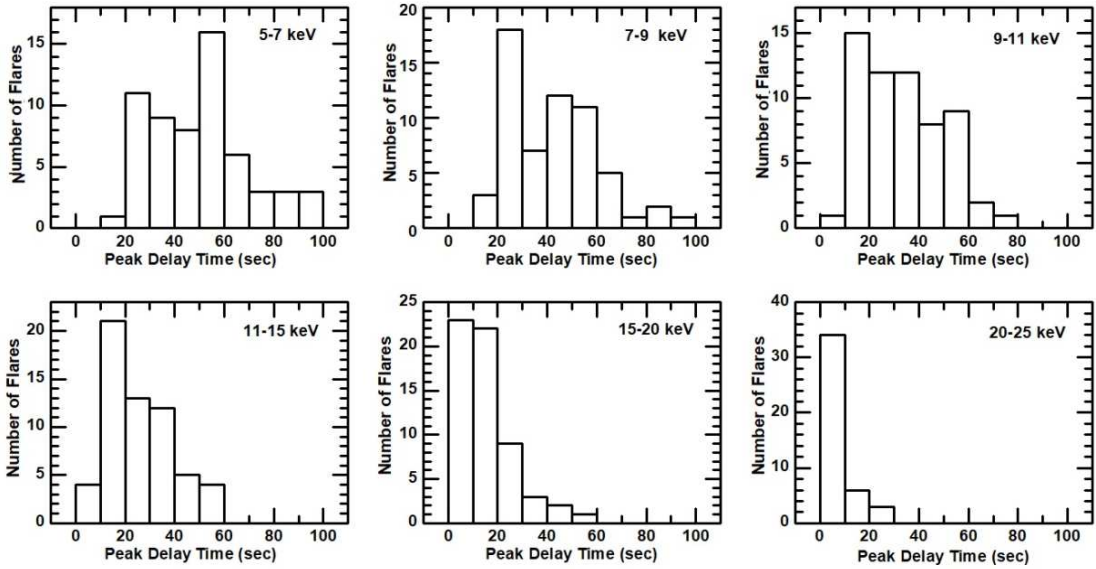

Fig. 10: Histogram of delay time corresponding to peak 1 by considering 25-30 keV as a reference band.

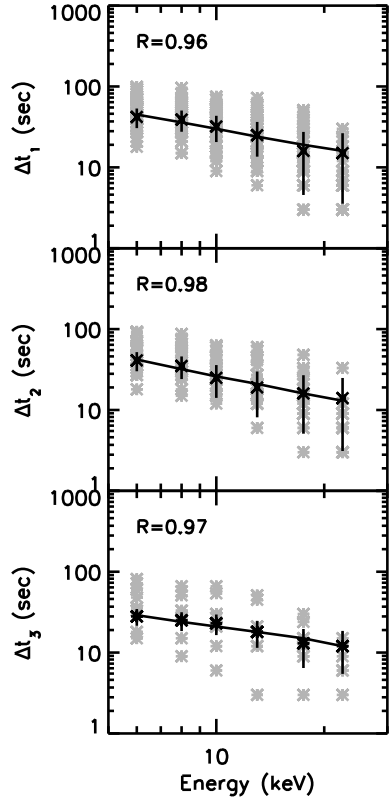

Fig. 11: Relative time delay of the peak time in a given energy band with respect to the reference energy band (25-30 keV) as a function of energy with different number of peaks, respectively. Correlation coefficient is determined by fitting the binned data points (black symbol) while the original data points are also plotted in grey for the reference.

We calculate the peak plasma temperature (T) and the differential emission measure (DEM) by minimizing $\chi^2$ between the model and observed SOXS X-ray spectra. It is to note that, we have discarded spectra with pulse pile-up to avoid spurious results. We have selected a representative set of 14 flares for spectra analysis purpose in this investigation. Figure 12 (a) and (b) show spectral evolution of the flare occurred on 14 Aug. 2004 and 17 Sept. 2005, respectively in the energy band of 4-25 keV of the Si detector. The spectral fit is made employing multi-thermal and single power-law functions provided in OSPEX. For multi-thermal model, we made use of 'multi_therm_pow' function while '1pow' is deployed to fit the non-thermal component of the flare emission. The multi-thermal model, used in the present study, employs power-law functional form of differential emission measure (DEM[T]) distribution over temperature (see Awasthi et al. (2016) for a comparative survey of various DEM[T] inversion techniques). The output parameters of the multi-thermal function include Power-law index of the DEM[T] function as well as minimum and maximum temperature of the plasma, which have been subjected to vary as a free parameter during the iteration in order to derive the best-fitting to the observed spectra. On the other hand, the power-law function employed to fit the non-thermal fraction of the X-ray emission, gives the output parameters such as normalization value as well as power-law index of the spectral shape in high energy. The modelled flux contribution from the multi-thermal function is shown by the yellow line and that of the single power-law function is shown by the green line. The total of these two contributions is shown by the red line. The total modelled flux is a good fit over the observed spectra (black) with the goodness of fit evaluated in terms of reduced $\chi^2$.

For the spectral fit corresponding to SOXS spectra recorded during 05:42:48-05:43:45 UT for 14 Aug 2004 flare, we obtained the DEM=$0.311 \times 10^{48}$ cm$^{-3}$ keV$^{-1}$, maximum temperature =1.22 keV (14.15 MK) and a DEM power-law index $\delta$=6.18 while the $\chi^2$ value of the fit reached as good as 2.17. Similar analysis for all the spectra prepared during entire flare is made which enables us to make a temporal evolution analysis of all the spectral fit parameters as shown in figure 13 for the cases of 14-Aug-2004 and 17-Sep-2005 flares. From all the spectra fitted for 14 August 2004 event, we estimate peak DEM=$3.36 \times 10^{49}$ cm$^{-3}$ keV$^{-1}$ (at T=2 keV (23.2 MK)), peak temperature T=20.39 MK and a DEM power-law index $\delta$=7.51. Similarly for 17 September 2005 event ($\chi^2$=2.18), we estimate peak DEM=$0.14 \times 10^{49}$ cm$^{-3}$ keV$^{-1}$ (at T=2 keV), peak temperature T=27.09 MK and a DEM power-law index $\delta$=6.72. The peak negative power-law index ($\gamma$) derived from the single power-law fit for the above two events is 5.95 and 5.00 respectively. The physical significance of negative power-law index of observed photon spectra is directly linked with the spectral index of non-thermal electron electron beam Awasthi et al. (2014) and therefore peak value of $\gamma$ may be linked with the energetics of non-thermal processes. The same method is employed for 14 flare events under investigation and for each analyzed spectra for a given flare event we measured flare plasma parameters viz. DEM, T, $\delta$ and negative power-law index ($\gamma$) derived from single power-law fit. In Figure 13 we show the results of measurements of the flare plasma parameters and their variation over time of the flare. These measurements enabled us to study in greater detail the variation of flare plasma parameters as a function of time. In particular, we measure the time delay of DEM peak relative to temperature (T) peak. The peak DEM, peak T and delay time, and peak $\delta$ and $\gamma$ derived for each flare event are listed in Table III.

Table 3: Spectral characteristics of flare plasma

| Sr. No. | Date | Temperature | | DEM | | Delay Time (sec) | Peak Flux 4.0-25 (keV) [1] | Power law index($\delta$) |
|---|---|---|---|---|---|---|---|---|
| | | Peak Time (UT) | $T_{Peak}$ (Mk) | Peak Time (UT) | $DEM_{Peak}$ [2] | | | |
| 1 | 2004 Jan. 06 | 6:25:00 | 26.64 | 6:30:00 | 1.76 | 300 | 9.93E+06 | 5.09 |
| 2 | 2004 Mar. 25 | 4:37:00 | 28.04 | 4:40:00 | 0.22 | 180 | 1.16E+07 | 5.28 |
| 3 | 2004 Apr. 25 | 5:36:00 | 18.42 | 5:38:00 | 1.17 | 120 | 1.16E+07 | 5.12 |
| 4 | 2004 Jul. 14 | 5:22:00 | 18.86 | 5:24:00 | 2.18 | 120 | 3.21E+07 | 5.02 |
| 5 | 2004 Aug. 14 | 5:43:00 | 20.39 | 5:44:00 | 3.36 | 60 | 6.76E+07 | 7.51 |
| 6 | 2004 Oct. 30 | 6:18:00 | 18.87 | 6:20:00 | 1.98 | 120 | 3.17E+07 | 5.84 |
| 7 | 2004 Oct. 31 | 5:32:00 | 30.01 | 5:36:00 | 1.23 | 240 | 1.38E+07 | 5.45 |
| 8 | 2005 Jan. 15 | 4:31:00 | 20.76 | 4:32:00 | 2.2 | 60 | 8.69E+07 | 6.26 |
| 9 | 2005 Jul. 27 | 4:59:00 | 18.42 | 5:03:00 | 2.04 | 240 | 1.81E+07 | 5.62 |
| 10 | 2005 Aug. 03 | 5:03:00 | 31.52 | 5:09:00 | 0.8 | 360 | 1.57E+07 | 5.7 |
| 11 | 2005 Aug. 25 | 4:40:00 | 38.52 | 4:42:00 | 0.23 | 120 | 3.81E+07 | 6.41 |
| 12 | 2005 Sep. 09 | 5:39:00 | 17.88 | 5:42:00 | 2.78 | 180 | 3.90E+07 | 5.84 |
| 13 | 2005 Sep. 10 | 6:11:00 | 18.54 | 6:12:00 | 0.88 | 60 | 2.36E+07 | 6.2 |
| 14 | 2005 Sep. 17 | 6:07:00 | 27.09 | 6:09:00 | 0.14 | 120 | 5.15E+07 | 6.72 |

It may be noted from Figure 13 that T peaks earlier than DEM peak and the time difference varies in the range of 60 and 360 s in the flare events. We measured the time (seconds) by which DEM peak delays to the peak time of temperature T for each of 14 flares.

Next, as shown in Figure 14 is the delay time ($\Delta t$) of the DEM peak relative to peak of T plotted as a function of peak flare flux in 4-25 keV recorded for the flares under investigation. We report, for the first time that this delay time reduces with the increase in the peak flux and found to be ∼100 s for the high-intensity class flares while amounts to ∼300 s for the low intensity flares. This suggests that hot plasma content increases faster in higher magnitude flares than that in the low intensity class flares.

## 5 DISCUSSIONS AND CONCLUSIONS

Temporal study of 60 flares in the energy range 5-30 keV using Si and CZT detector of the SOXS mission explicitly reveals that time evolution of X-ray emission during the flares have growth in 1-3 steps. In our investigation, we could not find the flares with more than three peaks, which, however may be limited to the choice of energy band for determining the number of peaks, which in our case is low-energy X-ray emission. As the SXR emission is a cumulative effect of the non-thermal processes giving rise to the hard X-ray emission (Dennis & Zarro 1993), a few less-dominant transients (peaks) in HXR flux may not appear in SXR. In agreement, HXR emission is found to depict more bursty nature with multiple peaks as investigated by Grigis & Benz (2005). Our study has shown that almost 50% of the flares (29 flares) have one peak, while 23 flares have double peak evolution, and only 8 flares reveal 3 steps of maximum. The above classification does not depend on the flare magnitude as in each category we have M and C-class flares. Similarly, we also found that the rise time of the flare does not depend significantly upon the flare magnitude (peak flux), which is in agreement to earlier investigation made by Veronig et al. (2002).

---

[1] Peak Flux unit: Photons s$^{-1}$ cm$^{-2}$ keV$^{-1}$

[2] Peak DEM unit is $10^{49}$ cm$^{-3}$ keV$^{-1}$.

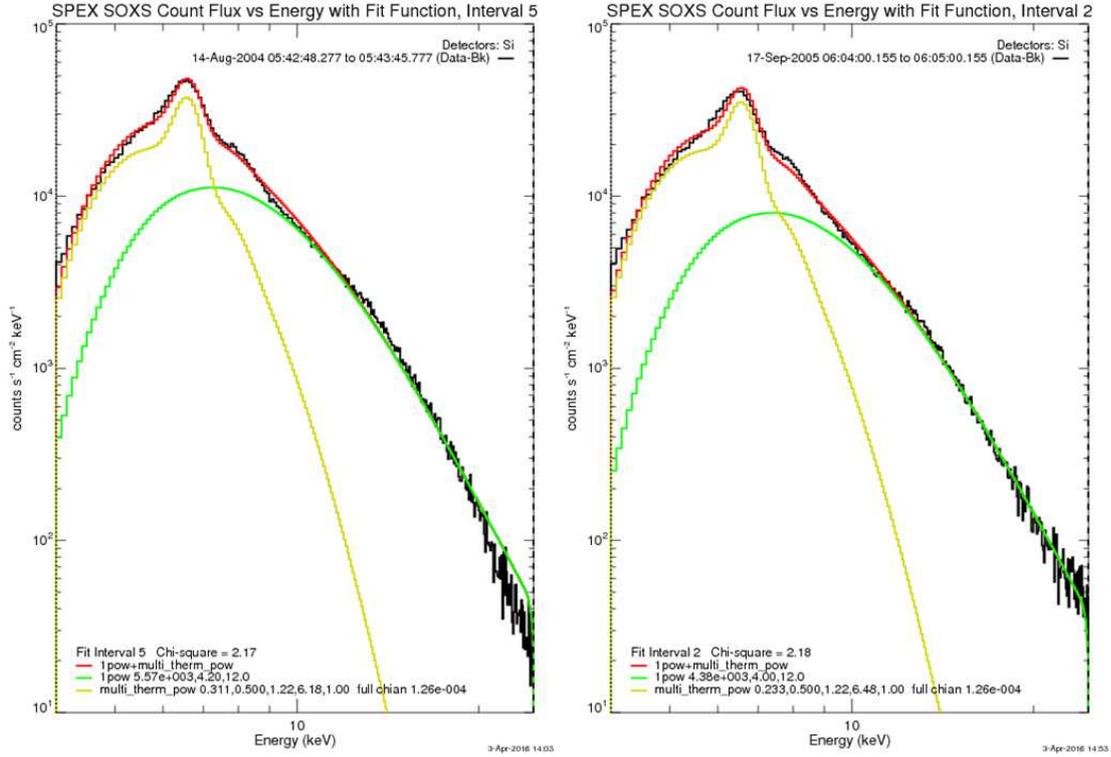

Fig. 12: Spectral fit of the 14 August 2004 (left panel) and 17 September 2005 (right panel) flare events in the energy range of 4-25 keV as observed by the Si detector onboard the SOXS mission (shown by black line). The spectral fit with the help of multi-thermal and single power-law functions is shown respectively by yellow and green lines. The goodness of fit of the total modelled flux (red line) over the observed spectra is evaluated to be $\chi^2$=2.17 (left panel) and 2.18 (right panel), respectively.

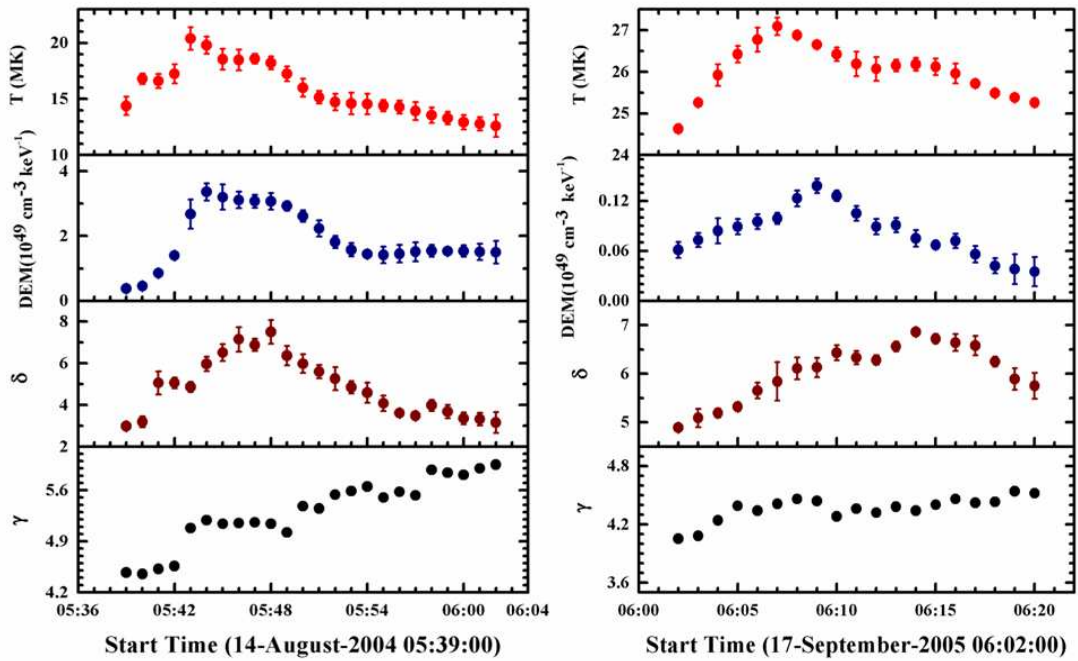

Fig. 13: Variation of the flare plasma parameters: T, DEM, $\delta$ and $\gamma$ as a function of time.

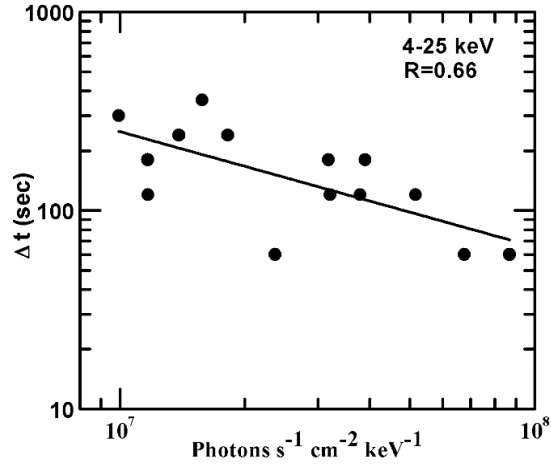

Fig. 14: Delay time (Δt) variation as a function of peak flare flux in 4-25 keV.

It appears that the flare evolution (rise time and peak flux) is dictated by the magnetic field configuration of the active region in which the flare occurs as shown in rise time and peak magnitude of the flare correlation with the magnetic complexity index, $M_j$. This figure reveals unambiguously key role of magnetic field complexity in governing the magnitude and rise time of the flare. Further, we find strong correlation (R=0.83) between the peak flux in soft (5-7 keV) and hard (25-30 keV) X-ray bands. This suggests that, although delayed in time, the peak magnitude of SXR emission produced as a consequence of the thermal-bremsstrahlung of the chromospheric evaporated plasma, is directly linked with the electrons accelerated during reconnection process producing HXR emission in a direct interaction with the chromosphere. Moreover, the time derivative of soft X- ray emission tend to mimic temporal evolution of hard X-ray emission known as Neupert effect as shown in present investigation.

The mean rise time of the flares estimated by Veronig et al. (2002) in 1.55-12.4 keV energy band is 546 s. On the other hand, we obtain mean rise time 433 and 147 s in 5-7 and 25-30 keV respectively. This motivated us to measure the rise time as a function of energy bands and it revealed that rise time decreases as a function of energy, which is in agreement to Jain et al. (2011b). The decreasing trend of rise time as a function of increasing energy is a direct consequence of energy-dependent cooling time. For higher energies, the cooling time scale is lesser in compared to the lower energy X-ray emission as determined from the analysis of 10 M-class flares observed by SOXS, Jain et al. (2011b) who probed the role of energy-dependent multi-thermal delay in the evolution of X-ray emission. On the other hand, our analysis considers larger number of flares, and with better coverage of flare intensity class than that analyzed in Jain et al. (2011b). During the rise phase of the flare, conduction cooling dominates over the radiative cooling. In this regard, the rise time of the flare will be affected by conduction cooling.

Our current investigation employing spectra analysis of the 14 flares observed by Si detector (4-25 keV) reveals that peak plasma temperature of the flare reaches to maximum earlier than the maximum of the differential emission measure (DEM) of the flare by 60-360 s. The temperature in a flare increases as a consequence of thermalization of the ambient flare plasma due to electron bremsstrahlung in the rise phase. The increased temperature further ionizes the flare plasma, which, in turn, enhances the electron density in a given flare volume Jain et al. (2006, 2011b). We report, for the first time, the delay time ($\Delta t$) of the DEM-peak relative to the temperature-peak reduces with the increase in the peak flux in 4-25 keV. This may be quantified in terms of higher chromospheric evaporation rate for large magnitude flares compared with that resulting for low-magnitude flares. The faster chromospheric evaporation leads to rapid filling of the coronal loop and therefore appearing in the form of less-delayed from the temperature peak.

Further, the statistical relationship between rise time and peak flux reveals weak correlation. On the contrary, rise time as well as peak flux are found to be dependent on the magnetic complexity of the active region. The detailed spectral analysis of 14 flares revealed that peak plasma temperature of the flare reaches to maximum earlier than the maximum of the DEM of the flare by 60-360 s. However, the delay time of DEM-peak with respect to peak of plasma flare temperature decreases with peak flux of the flare. This suggests that high magnitude (flux) flares have higher plasma temperature and consequently causes significantly higher ionization with faster rate so as to enhance the electron density and thereby emission measure.

**Acknowledgements** This research work has been accomplished with the grant received from Gujarat Council on Science and Technology (GUJCOST), Dept. of Science & Technology, Govt. of Gujarat under minor research project grants scheme. We express sincere thanks to GUJCOST for the financial assistance. We also express our gratitude to Sh. Vallabh M. Patel, President of Kadi Sarva Vishwavidyalaya for providing full support for research. The SOXS mission was launched by ISRO and the data is linked at Physical Research Laboratory server. AKA is supported by NSFC 41474151, 41774150, and 4171101125 and the International postdoctoral program of USTC. We sincerely thank the anonymous referee for constructive comments which improved the scientific clarity of the work.

Appendix Table 1: Physical Properties of solar flares under current investigation

| Sr. No. | Date | 9-11 keV | | | 15-20 keV | | | 25-30 keV | | |
|---|---|---|---|---|---|---|---|---|---|---|
| | | Start Time (U.T.) | Peak Time (U.T.) | End Time (U.T.) | Start Time (U.T.) | Peak Time (U.T.) | End Time (U.T.) | Start Time (U.T.) | Peak Time (U.T.) | End Time (U.T.) |
| 1 | 2003 Jul. 30 | 04:07:33 | 04:09:36 | 04:20:36 | 04:07:48 | 04:09:27 | 04:50:03 | 04:08:00 | 04:09:09 | 04:50:03 |
| 2 | 2003 Nov. 13 | 04:59:14 | 05:01:59 | 05:10:44 | 04:59:41 | 05:01:47 | 05:08:14 | 05:00:05 | 05:01:20 | 05:05:56 |
| 3 | 2003 Nov. 19 | 03:57:26 | 04:00:17 | 04:19:02 | 03:57:41 | 04:00:02 | 04:37:56 | 03:57:59 | 03:59:35 | 04:40:44 |
| 4 | 2004 Jan. 06 | 06:17:20 | 06:22:17 | 06:42:47 | 06:17:47 | 06:21:44 | 06:37:08 | 06:18:11 | 06:21:41 | 06:37:02 |
| 5 | 2004 Jan. 07 | 03:56:00 | 04:00:57 | 04:29:06 | 03:56:00 | 04:00:54 | 05:31:21 | 03:56:18 | 04:00:03 | 05:31:30 |
| 6 | 2004 Jan. 08 | 04:56:12 | 05:04:33 | 05:20:00 | 04:56:54 | 05:04:12 | 05:14:03 | 04:56:54 | 05:04:12 | 05:14:03 |
| 7 | 2004 Jan. 10 | 04:18:39 | 04:20:03 | 04:35:54 | 04:19:06 | 04:19:51 | 04:51:18 | 04:19:33 | 04:19:42 | 04:41:51 |
| 8 | 2004 Jan. 10 | 05:10:27 | 05:12:54 | 05:24:30 | 05:10:45 | 05:12:45 | 05:21:24 | 05:11:09 | 05:12:30 | 05:21:42 |
| 9 | 2004 Jan. 19 | 04:52:14 | 04:55:11 | 05:03:14 | 04:52:38 | 04:54:50 | 05:08:50 | 04:53:23 | 04:54:38 | 05:03:53 |
| 10 | 2004 Feb. 09 | 05:51:47 | 05:54:02 | 06:06:38 | 05:51:56 | 05:53:17 | 06:37:53 | 05:52:14 | 05:53:08 | 06:38:59 |
| 11 | 2004 Mar. 22 | 06:10:04 | 06:15:34 | 06:35:52 | 06:10:49 | 06:14:40 | 07:01:34 | 06:11:01 | 06:14:25 | 07:01:34 |
| 12 | 2004 Mar. 25 | 04:29:45 | 04:33:54 | 05:04:15 | 04:30:24 | 04:33:48 | 05:58:30 | 04:30:57 | 04:33:33 | 05:58:30 |
| 13 | 2004 Apr. 05 | 05:36:11 | 05:43:14 | 06:30:05 | 05:37:11 | 05:42:50 | 06:30:05 | 05:38:47 | 05:42:17 | 06:27:20 |
| 14 | 2004 Apr. 11 | 03:58:59 | 04:10:14 | 04:45:35 | 04:01:14 | 04:10:05 | 05:40:35 | 04:01:14 | 04:10:05 | 05:40:35 |
| 15 | 2004 Apr. 25 | 05:29:27 | 05:32:54 | 05:54:27 | 05:29:57 | 05:32:27 | 05:51:21 | 05:30:15 | 05:32:18 | 05:47:33 |
| 16 | 2004 Jul. 12 | 04:29:18 | 04:32:03 | 04:54:06 | 04:30:00 | 04:31:27 | 05:14:30 | 04:31:06 | 04:31:12 | 05:13:48 |
| 17 | 2004 Jul. 13 | 05:26:40 | 05:29:31 | 05:34:25 | 05:27:04 | 05:28:46 | 05:34:25 | 05:28:04 | 05:28:16 | 05:34:25 |
| 18 | 2004 Jul. 14 | 05:17:31 | 05:18:49 | 05:33:22 | 05:17:46 | 05:18:31 | 05:33:22 | 05:17:52 | 05:18:25 | 05:33:22 |
| 19 | 2004 Jul. 21 | 05:12:26 | 05:17:23 | 05:29:23 | 05:12:59 | 05:17:08 | 05:27:32 | 05:13:20 | 05:16:58 | 05:29:35 |
| 20 | 2004 Aug. 14 | 04:12:06 | 04:14:15 | 04:45:15 | 04:12:15 | 04:14:09 | 04:55:12 | 04:12:24 | 04:14:00 | 04:54:18 |
| 21 | 2004 Aug. 14 | 05:37:06 | 05:40:27 | 06:03:57 | 05:38:18 | 05:40:21 | 06:03:57 | 05:38:33 | 05:40:06 | 06:03:57 |
| 22 | 2004 Aug. 17 | 05:00:40 | 05:02:28 | 05:27:46 | 05:01:01 | 05:02:19 | 05:40:25 | 05:01:10 | 05:02:10 | 05:33:43 |
| 23 | 2004 Aug. 31 | 05:29:08 | 05:35:11 | 05:49:38 | 05:29:44 | 05:34:41 | 06:04:26 | 05:29:44 | 05:34:41 | 06:04:26 |
| 24 | 2004 Oct. 28 | 06:03:39 | 06:04:45 | 06:11:03 | 06:03:48 | 06:04:42 | 06:11:09 | 06:03:57 | 06:04:33 | 06:10:27 |
| 25 | 2004 Oct. 30 | 06:10:57 | 06:13:48 | 06:24:21 | 06:11:03 | 06:13:33 | 06:24:21 | 06:11:18 | 06:13:15 | 06:24:21 |
| 26 | 2004 Oct. 31 | 05:25:18 | 05:29:54 | 05:50:06 | 05:25:33 | 05:29:21 | 06:04:06 | 05:26:00 | 05:29:12 | 06:06:27 |
| 27 | 2004 Nov. 07 | 04:12:48 | 04:14:45 | 04:28:27 | 04:12:51 | 04:13:57 | 04:29:51 | 04:13:00 | 04:13:48 | 04:28:18 |
| 28 | 2004 Nov. 07 | 04:45:54 | 04:47:33 | 04:58:39 | 04:46:03 | 04:47:45 | 04:57:18 | 04:46:15 | 04:46:57 | 05:10:51 |
| 29 | 2004 Nov. 19 | 05:06:02 | 05:09:50 | 05:19:53 | 05:06:23 | 05:09:35 | 05:15:59 | 05:08:11 | 05:09:05 | 05:15:20 |
| 30 | 2005 Jan. 15 | 04:10:37 | 04:13:58 | 04:24:25 | 04:11:40 | 04:13:52 | 04:23:49 | 04:12:13 | 04:13:40 | 04:23:25 |
| 31 | 2005 Jan. 15 | 04:29:10 | 04:30:34 | 04:43:07 | 04:29:16 | 04:30:31 | 05:00:55 | 04:29:34 | 04:30:16 | 05:00:55 |
| 32 | 2005 Jan. 21 | 04:19:57 | 04:25:24 | 04:45:48 | 04:20:06 | 04:25:15 | 05:04:09 | 04:20:06 | 04:25:15 | 05:04:09 |
| 33 | 2005 Jul. 01 | 04:58:16 | 05:00:46 | 05:06:58 | 04:58:31 | 05:00:07 | 05:05:01 | 04:58:31 | 05:00:07 | 05:05:01 |
| 34 | 2005 Jul. 03 | 04:49:36 | 04:51:09 | 05:00:09 | 04:50:06 | 04:50:48 | 04:55:00 | 04:50:06 | 04:50:48 | 04:55:00 |
| 35 | 2005 Jul. 27 | 04:36:50 | 04:47:14 | 05:28:47 | 04:38:02 | 04:46:56 | 05:28:47 | 04:38:53 | 04:46:47 | 05:28:47 |
| 36 | 2005 Jul. 30 | 05:08:02 | 05:12:47 | 05:39:17 | 05:08:32 | 05:12:14 | 05:43:23 | 05:08:32 | 05:12:14 | 05:43:23 |
| 37 | 2005 Aug. 03 | 04:57:03 | 05:01:48 | 05:20:15 | 04:57:27 | 05:01:42 | 05:41:48 | 04:58:03 | 05:01:30 | 05:42:18 |
| 38 | 2005 Aug. 25 | 04:34:24 | 04:39:12 | 04:57:24 | 04:34:48 | 04:39:06 | 05:19:21 | 04:35:18 | 04:38:54 | 05:19:18 |
| 39 | 2005 Sep. 09 | 05:31:39 | 05:39:06 | 06:05:39 | 05:31:48 | 05:38:45 | 06:05:12 | 05:32:09 | 05:38:39 | 06:04:18 |
| 40 | 2005 Sep. 12 | 04:47:57 | 05:00:39 | 05:21:27 | 04:48:15 | 05:00:24 | 05:48:18 | 04:48:15 | 05:00:24 | 05:48:18 |
| 41 | 2005 Sep. 17 | 05:57:03 | 06:00:48 | 06:20:15 | 05:57:39 | 06:00:45 | 06:41:48 | 05:58:00 | 06:00:36 | 06:42:18 |
| 42 | 2005 Nov. 14 | 03:57:27 | 03:58:36 | 04:06:15 | 03:57:39 | 03:58:33 | 04:05:42 | 03:57:45 | 03:58:18 | 04:12:57 |
| 43 | 2005 Nov. 14 | 04:17:27 | 04:21:21 | 04:34:39 | 04:17:36 | 04:20:48 | 04:55:21 | 04:18:03 | 04:20:21 | 04:32:39 |
| 44 | 2005 Dec. 01 | 04:52:14 | 04:53:47 | 05:11:11 | 04:52:41 | 04:53:35 | 05:04:17 | 04:52:41 | 04:53:35 | 05:04:17 |
| 45 | 2006 Apr. 06 | 05:26:30 | 05:32:39 | 05:42:54 | 05:26:45 | 05:32:18 | 06:18:06 | 05:28:00 | 05:32:03 | 06:11:36 |

| Sr. No. | Date | 9-11 keV | | | 15-20 keV | | | 25-30 keV | | |
|---|---|---|---|---|---|---|---|---|---|---|
| | | Start Time (U.T.) | Peak Time (U.T.) | End Time (U.T.) | Start Time (U.T.) | Peak Time (U.T.) | End Time (U.T.) | Start Time (U.T.) | Peak Time (U.T.) | End Time (U.T.) |
| 46 | 2006 Dec. 05 | 05:03:54 | 05:07:15 | 05:13:18 | 05:04:24 | 05:06:51 | 05:13:30 | 05:04:24 | 05:06:51 | 05:13:30 |
| 47 | 2006 Dec. 07 | 04:33:17 | 04:37:35 | 04:49:14 | 04:33:29 | 04:37:26 | 04:41:05 | 04:33:29 | 04:37:26 | 04:41:05 |
| 48 | 2010 Feb. 08 | 05:14:51 | 05:21:39 | 05:31:18 | 05:15:09 | 05:21:30 | 05:29:00 | 05:15:18 | 05:21:06 | 05:32:36 |
| 49 | 2010 Oct. 31 | 04:27:40 | 04:28:40 | 04:39:04 | 04:27:43 | 04:28:34 | 05:05:19 | 04:27:58 | 04:28:28 | 05:04:52 |
| 50 | 2010 Nov. 06 | 04:40:21 | 04:46:18 | 04:58:57 | 04:41:45 | 04:46:12 | 05:31:00 | 04:41:45 | 04:46:12 | 05:31:00 |
| 51 | 2011 Feb. 14 | 04:29:34 | 04:37:28 | 05:17:34 | 04:30:55 | 04:36:37 | 05:09:58 | 04:31:25 | 04:36:34 | 05:06:13 |
| 52 | 2011 Feb. 15 | 04:28:49 | 04:30:01 | 04:37:10 | 04:29:13 | 04:29:58 | 04:36:01 | 04:29:13 | 04:29:58 | 04:36:01 |
| 53 | 2011 Feb. 16 | 05:42:09 | 05:44:48 | 05:54:06 | 05:42:18 | 05:44:30 | 05:59:27 | 05:42:51 | 05:44:06 | 05:55:12 |
| 54 | 2011 Feb. 18 | 04:46:58 | 04:49:22 | 05:05:07 | 04:47:25 | 04:49:19 | 05:29:22 | 04:47:25 | 04:49:19 | 05:29:22 |
| 55 | 2011 Mar. 07 | 05:02:27 | 05:08:06 | 05:21:30 | 05:03:18 | 05:07:39 | 05:14:00 | 05:03:18 | 05:07:39 | 05:14:00 |
| 56 | 2011 Mar. 09 | 04:58:52 | 05:01:52 | 05:05:28 | 04:58:58 | 05:01:34 | 05:06:52 | 04:58:58 | 05:01:34 | 05:06:52 |
| 57 | 2011 Mar. 11 | 04:29:11 | 04:31:41 | 04:36:11 | 04:29:23 | 04:31:35 | 04:36:59 | 04:29:41 | 04:31:23 | 04:37:08 |
| 58 | 2011 Mar. 12 | 04:35:39 | 04:39:36 | 04:49:21 | 04:35:57 | 04:39:21 | 05:00:57 | 04:36:27 | 04:39:03 | 05:00:57 |
| 59 | 2011 Apr. 14 | 05:23:29 | 05:24:38 | 05:31:50 | 05:23:47 | 05:24:11 | 05:34:50 | 05:23:47 | 05:24:11 | 05:34:50 |
| 60 | 2011 Apr. 22 | 04:37:40 | 04:40:34 | 05:17:49 | 04:37:52 | 04:40:13 | 05:10:49 | 04:38:28 | 04:40:01 | 05:09:37 |